\documentclass[12pt,preprint]{aastex}
\usepackage{epsfig}
\shorttitle{Interacting Binary BD$+36^\circ 4063$}
\shortauthors{Williams et al.}

\begin{document}


\title{The Evolutionary State of the Massive Interacting Binary 
 BD+36$^\mathbf{\circ}$4063} 

\author{S. J. Williams\altaffilmark{1}, D. R. Gies, 
  R. A. Matson\altaffilmark{1}}
\affil{Center for High Angular Resolution Astronomy and
 Department of Physics and Astronomy,
 Georgia State University, P. O. Box 4106, Atlanta, GA 30302-4106; 
 swilliams@chara.gsu.edu, gies@chara.gsu.edu, rmatson@chara.gsu.edu}

\altaffiltext{1}{Visiting Astronomer, Kitt Peak National Observatory,
National Optical Astronomy Observatory, operated by the Association
of Universities for Research in Astronomy, Inc., under contract with
the National Science Foundation.}

\author{W. Huang\altaffilmark{1}}
\affil{Department of Astronomy, University of Washington, 
Box 351580, Seattle, WA 98195-1580; hwenjin@astro.washington.edu }


\begin{abstract}
We present a spectroscopic and photometric analysis of the 
remarkable massive binary system, BD$+36^\circ 4063$.  
We argue that the visible ON star is undergoing a rapid
mass transfer episode that results in a thick and opaque
disk that surrounds and renders invisible its massive companion. 
A comparison of the projected rotational velocity 
and the orbital semiamplitude of the visible star indicates
a mass ratio near unity.  Models for conservative mass 
transfer show that the equal mass state occurs at the 
point of minimum separation, and we argue that BD$+36^\circ 4063$
may represent the first system identified at this rapid 
and rare stage of massive binary evolution. 
\end{abstract}
\keywords{binaries: spectroscopic  --- 
stars: early-type  ---
stars: evolution ---
stars: individual (BD$+36^\circ 4063$)}


\setcounter{footnote}{1}

\section{Introduction}                              

Massive close binaries are destined to undergo extreme transformations
once they begin to exchange mass and angular momentum \citep{lan08}. 
If mass transfer is conservative, then the system separation 
decreases as mass transfer progresses, and the orbit dimensions 
reach a minimum once a mass ratio of unity is reached.  
The system then experiences a critical phase where 
the mass transfer rates may exceed $10^{-4} M_\odot$~y$^{-1}$ \citep{wel01}. 
Subsequently, the orbit will widen after the mass donor becomes the 
lower mass object, and a slower and more extended mass transfer stage will 
begin.  The gainer accepts both mass and angular momentum, and it 
may quickly reach a state of critical rotation where gravitational 
and centripetal accelerations balance at the equator.  This may limit 
further mass accretion and lead to the formation of a thick, opaque, and 
mass losing disk surrounding the gainer star \citep{naz06}.

This stage of binary evolution is probably represented by 
the W~Serpentis class of massive binaries \citep{pla80,tar00}.  
These binaries contain a visible, low gravity,  
donor star that fills its Roche lobe,
while the higher mass, gainer star is invisible in the composite spectrum
because it is immersed in a thick disk.  The first direct image 
of such a disk around a gainer star was recently made with the
CHARA Array for the best known member of the class $\beta$~Lyrae 
\citep{zha08}.  The shortest period W~Ser binary known is RY~Scuti
($P=11.1$~d).  It consists of a $7 M_\odot$, O9.7~Ibpe 
star and a $28 M_\odot$ hidden companion \citep{gru07}, 
and the entire system is surrounded by a 2000~AU wide gas and dust 
torus \citep{smi02}.  The known W~Ser systems are probably 
experiencing the slow mass transfer stage after
mass ratio reversal, and no W~Ser system has yet been identified at the 
close, rapid mass transfer stage. 

The missing link may be filled by the subject of this paper, BD$+36^\circ4063$.
The star (ON9.7~Iab; \citealt{wal00}) is located 
in the Cyg~OB1 association \citep{hum78}. 
It was noted by \citet{mat89} as a key example of an ON star, 
showing spectral features indicating the N-enrichment and C-depletion 
characteristic of CNO-processed elements \citep{wal00}.  Many ON stars 
are members of binary systems \citep{bol78} where mass transfer may 
remove the H envelope and reveal CNO-processed gas in the atmosphere. 
Thus, it is very pertinent that BD$+36^\circ4063$ was discovered by 
Howarth \citep{wal00,har02} to be a single-lined, spectroscopic binary 
with a period of 4.8~d.  In a recent meeting contribution\footnote{
http://www.lowell.edu/workshops/Contifest/talks/Howarth.pdf},
Howarth presents radial velocity and light curves, and he shows that 
although the companion is massive, its absorption features 
are completely absent from the observed spectrum. 

We obtained blue spectra of this and several other 
faint O-stars in a search for binaries, unaware at the time of 
Howarth's detection of binary motion.  Here we confirm the orbital period 
and present preliminary orbital elements for the binary.  We argue
that the visible star is filling its Roche lobe and that the high rate of 
mass transfer results in a dense disk that blocks a direct view of 
the massive companion.  We conclude with a brief discussion of system 
parameters and evolutionary status.


\section{Observations and Orbital Elements}         

Seven spectra of BD$+36^\circ4063$ were obtained with the Kitt Peak National
Observatory (KPNO) 2.1 m telescope from 2008 November 15 to 21.
These observations made use of the Goldcam spectrograph with
grating G47 (831 grooves mm$^{-1}$) in second order with a CuSO$_4$ 
order sorting filter.  The detector was the T3KC CCD 
(a $3072\times1024$ pixel array with $15\times 15$ $\mu$m pixels), 
and the resulting spectra have a resolving power of 
$R = \lambda / \Delta \lambda = 2400$ as measured from the
HeNeAr comparison lines.  Exposures were usually 600~s in duration, 
leading to spectra with a S/N = 200 per pixel in the continuum.  
The wavelength range is 3942 -- 5032 \AA ~with a
wavelength calibration accuracy of $\sim$5 km~s$^{-1}$ based on the rms 
scatter of fits to comparison lines and on the variance of multi-night 
measurements of the spectra of other stars.  
The spectra were extracted and calibrated using
standard routines in IRAF\footnote{IRAF is distributed by the National
Optical Astronomy Observatory, which is operated by the Association of
Universities for Research in Astronomy, Inc., under cooperative agreement
with the National Science Foundation.}, and then each 
continuum-rectified spectrum was transformed to a common 
heliocentric wavelength grid in $\log \lambda$ increments. 

Only one set of spectral lines was readily apparent in these spectra,
so we determined radial velocities by cross-correlating each spectrum
with a model spectrum from the 
TLUSTY/SYNSPEC BSTAR2006 grid \citep{lan07}.  We selected a model 
for $T_{\rm eff}= 28$ kK, $\log g = 3.0$, and microturbulent velocity 
of 10 km~s$^{-1}$, parameters that are typical for the star's classification
\citep{rya02,mar05}, although small variations in these stellar parameters do 
not significantly alter the resulting velocities.  
We selected a  N-enriched ``CN'' model to match better the line spectrum. 
The model spectrum was transformed to the observed grid by 
integration and was then convolved with functions 
to account for rotational broadening (\S3) and instrumental 
broadening.  A number of spectral regions that contain ISM features 
or emission lines were omitted from the cross-correlation sample (all hydrogen
lines, the diffuse interstellar band at $\lambda 4428$, and \ion{He}{2} 
$\lambda 4686$). Table~1 lists the date of observation, spectroscopic phase, 
radial velocity and error, and the observed minus calculated 
residual from the fit (below). 

\placetable{tab1}      

Our seven day time span is much too short to derive an accurate 
orbital period, so we estimated the period from the ellipsoidal variations 
in flux related to the tidal distortion of the star.  
Two sets of photometric measurements exist from 
all sky survey experiments.  The first set of 111 Cousins $I_C$
measurements were made between 2003 and 2007 by The Amateur Sky Survey 
(TASS\footnote{http://sallman.tass-survey.org/servlet/markiv/}; 
\citealt{dro06}).
The second set of 84 points were made in 1999 with the Northern Sky Variability
Survey (NSVS\footnote{http://skydot.lanl.gov/nsvs/nsvs.php}; 
\citealt{woz04}). Because the NSVS 
measurements record a broad spectral range (4500 -- 10000 \AA ),
we simply subtracted a constant value of 1.099 mag to bring
their mean into coincidence with the mean of the $I_C$-band
results. Despite differences in wavelength intervals, both the amplitude 
of variation and the ephemeri determined from each data set matched 
within uncertainties.
A discrete Fourier transform of this combined time 
series immediately showed evidence of half the orbital period. A double 
sinusoid light curve is exhibited in one orbit by tidally distorted stars, 
and for our data this yielded a period of
$P = 4.8126 \pm 0.0004$~d and an epoch of ON star superior 
conjunction at $T_{\rm SC}=$ HJD~2,452,448.60~$\pm~0.08$.

We then determined the remaining elements using a fit
of the velocities made with 
the non-linear, least-squares fitting program of \citet{mor74}. 
The weights were set to unity for most measurements (since they
have comparable measurement errors), but we increased the weight 
of the single datum in the negative branch to four (to balance
the four measurements in the positive branch) and we decreased 
the weight of the two conjunction phase measurements to 0.5.  
We fixed the period at the photometrically derived value  
and then solved for the circular orbital elements: 
systemic velocity $\gamma = -17 \pm 3$ km~s$^{-1}$, 
velocity semiamplitude $K = 163 \pm 3$ km~s$^{-1}$, and 
epoch of superior conjunction of the ON star 
$T_{\rm SC} = $ HJD~2,454,787.26~$\pm 0.03$.
Trial elliptical solutions did not improve the fit. 
The rms of the fit, 6.6 km~s$^{-1}$, is larger than 
the formal errors but comparable to what we find in observations 
of other O-stars from this run.  The spectroscopic epoch 
occurs $(0.3 \pm 0.2)$~d earlier than the prediction from the light 
curve ephemeris, which may indicate that the period is decreasing.
The possibility of a measurable period change offers us an important 
diagnostic of the mass transfer rate that must be confirmed in 
future observations. 


\section{Binary Properties}      

The most surprising feature of the orbital solution is that the 
mass function is large, $f(m)=(2.18 \pm 0.12) M_\odot$, 
suggesting that the companion is a massive star.  
However, there is no clear evidence of absorption lines of 
the companion in the individual spectra.  The expected radial 
velocities for any companion lines depend on the assumed mass ratio. 
We can estimate the mass ratio by considering the projected 
rotational velocity $V\sin i$ of the ON star.  If we assume 
that the ON star fills its Roche lobe and rotates synchronously, 
then the ratio of $V\sin i$ 
to semiamplitude $K$ is a monotonically increasing function 
of $Q= M_1/M_2$ (mass of the ON star divided by that of the hidden star;  
\citealt{gie86}) 
$${{V\sin i}\over K} = (Q+1) \Phi(Q)$$
where $\Phi(Q)$ is the fractional Roche radius of the ON star 
\citep{egg83}. 
Thus, we can use a measurement of $V\sin i$ to determine the mass ratio.
The resolution of our spectra is just adequate for 
this task.  We made measurements of 
the FWHM of the deep and relatively unblended line of 
\ion{Si}{4} $\lambda 4088$, whose profile should be dominated 
by rotational broadening.  The mean width of the line is 
FWHM = $2.95\pm 0.11$ \AA , which is significantly larger than 
the instrumental broadening measured in comparison lines 
near this wavelength, FWHM = $1.90 \pm 0.02$ \AA .  We then 
created synthetic spectra for a grid of test values of $V\sin i$ 
by convolving the model spectrum with a rotational broadening 
function \citep{gra05} for a linear limb darkening coefficient
of $\epsilon = 0.37$\citep{wad85}. It should be noted that varying the 
linear limb darkening coefficient even by a factor of two has no effect, 
within uncertainties, on $V\sin i$ measurements. These model profiles 
match the observed FWHM for $V\sin i =126 \pm 15$ km~s$^{-1}$.  
Although it is possible that some of the apparent line broadening 
is due to macroturbulence in the atmosphere, a test using a 
presumed macroturbulent broadening with $\xi = 50$ km~s$^{-1}$
(at the high end for similar supergiants; \citealt{rya02}) 
resulted in the same $V\sin i$ because the instrumental broadening 
is so much larger than the expected macroturbulent broadening. 
Then the relation above leads to a mass ratio $Q= 1.02 \pm 0.17$, 
i.e., the stars are about equal in mass. 

We see no evidence of absorption lines moving in anti-phase 
with a velocity comparable to that of the ON star.  However, Figure~1 
illustrates the orbital variations of the H$\beta$ line in which  
an emission component does appear to share the orbital 
motion of the companion.  We made a preliminary Doppler tomographic 
reconstruction of the two spectral components \citep{bag94} assuming the 
mass ratio given above, and we found evidence of similar, anti-phase
moving emission lines in H$\gamma$, \ion{He}{1} $\lambda\lambda 
4387,4471,4713,4921$, and \ion{He}{2} $\lambda 4686$. 
We suggest that these emission features form in a disk surrounding 
the gainer star. Significant emission lines also are observed at H$\alpha$
(see Howarth's conference poster) and in the near-IR at \ion{He}{2} 2.058
$\mu$m and Br$\gamma$ 2.166 $\mu$m \citep{han96,tam96}.
A comparison of the line depths in the reconstructed 
spectrum of the ON star with those in the model indicates that 
the continuum flux from the disk is faint, $F_2/F_1 \approx 0.1$. 
Note that the absorption part of H$\beta$ appears deeper at conjunctions. 
The same deepening was observed in other lines that strengthen 
in slightly cooler atmospheres (\ion{Si}{3}, \ion{O}{2}). 
Similar changes are found in RY~Scuti and are probably 
related to the tidal extension and gravity darkening of 
the donor star \citep{gru07}. 

\begin{figure}
\begin{center}
{\includegraphics[angle=0,height=12cm]{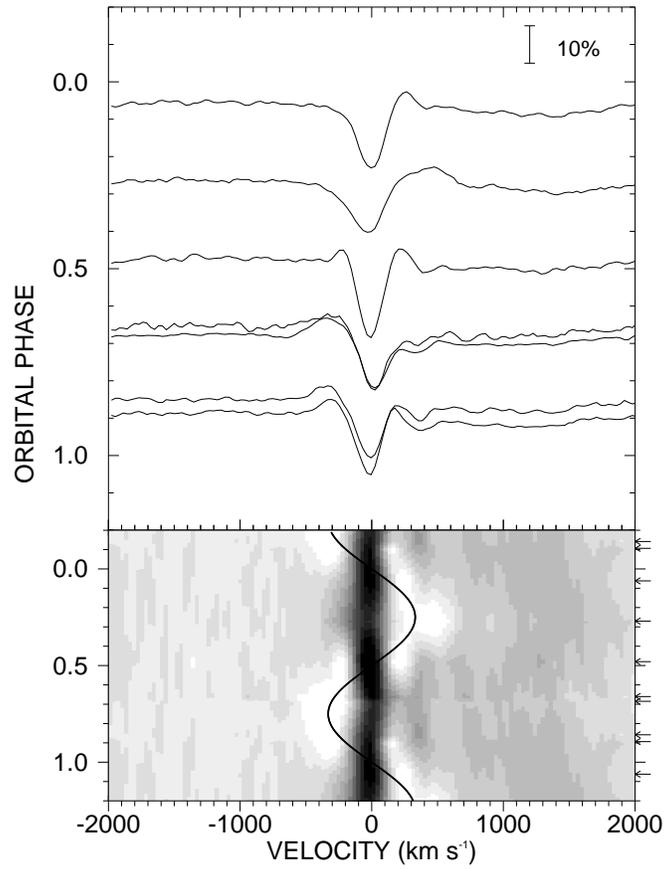}}
\end{center}
\caption{The variations of H$\beta$ $\lambda4861$ as a function
of orbital phase and velocity in the frame of the ON star.
The intensity between observed spectra in the gray-scale image 
is calculated by a linear interpolation between the closest 
observed phases (shown by arrows along the right axis).
The backwards S-curve in the gray-scale 
image shows the relative velocity curve of the companion star.
An emission component (bright in the gray-scale image)
appears to follow the velocity curve for the companion star.}
\label{fig1}
\end{figure}

\placefigure{fig1}     

The example of RY~Scuti shows that mass transfer to a thick disk can 
lead to systemic mass loss from the binary and the formation of circumbinary 
gas and dust structures \citep{smi02}.  There is some indication 
that an infrared excess from such circumbinary material is
present in the spectral energy distribution (SED) of 
BD$+36^\circ 4063$.   We show in Figure~2 the available 
flux measurements based upon Johnson $UBV$ \citep{hil56,col96},
Str\"{o}mgren \citep{cra75,gra98} TASS $V$, $I_C$, \citep{dro06,bes98}, 
and 2MASS $JHK_S$ magnitudes \citep{coh03,skr06}, plus mid-IR fluxes 
from the Spitzer/IRS post basic calibrated data 
archive\footnote{http://irsa.ipac.caltech.edu/applications/Spitzer/Spitzer/} 
\citep{hou04}.  We fit these observed fluxes with the BSTAR2006 flux model 
to determine the reddening \citep{fit99} $E(B-V) = 1.28 \pm 0.06$ mag and 
ratio of total-to-selective extinction $R_V = 3.50\pm 0.10$ (consistent 
with previous reddening estimates; \citealt{pat03}).  
The limb darkened, angular diameter of the ON star 
is $\theta_{LD} = 97 \pm 8$ $\mu$as (after correction 
for a companion flux contribution of $F_2 / F_1 = 0.1$). 
The resulting radius -- distance relationship is 
$R_{1}/R_{\odot} = (10.4 \pm 0.8)~d({\rm kpc})$.   
There does appear to be a flux excess in the SED at wavelengths 
$> 8 \mu$m that may result from systemic mass loss. Could stellar winds 
account for this excess? Because the BSTAR2006 grid does not take winds into
account, we also fit an SED from a CMFGEN model atmosphere \citep{hil98}
with similar stellar parameters and a wind. 
The absolute fluxes from the CMFGEN model were 
systematically $5\%$ higher than those from the BSTAR2006 SED over the optical
to near-IR range. However, there is no evidence of a marked IR excess near 
10 $\mu$m from the wind in the CMFGEN model SED. This suggests that the 
observed excess may result instead from binary mass loss. 

\begin{figure}
\begin{center}
{\includegraphics[angle=90,height=12cm]{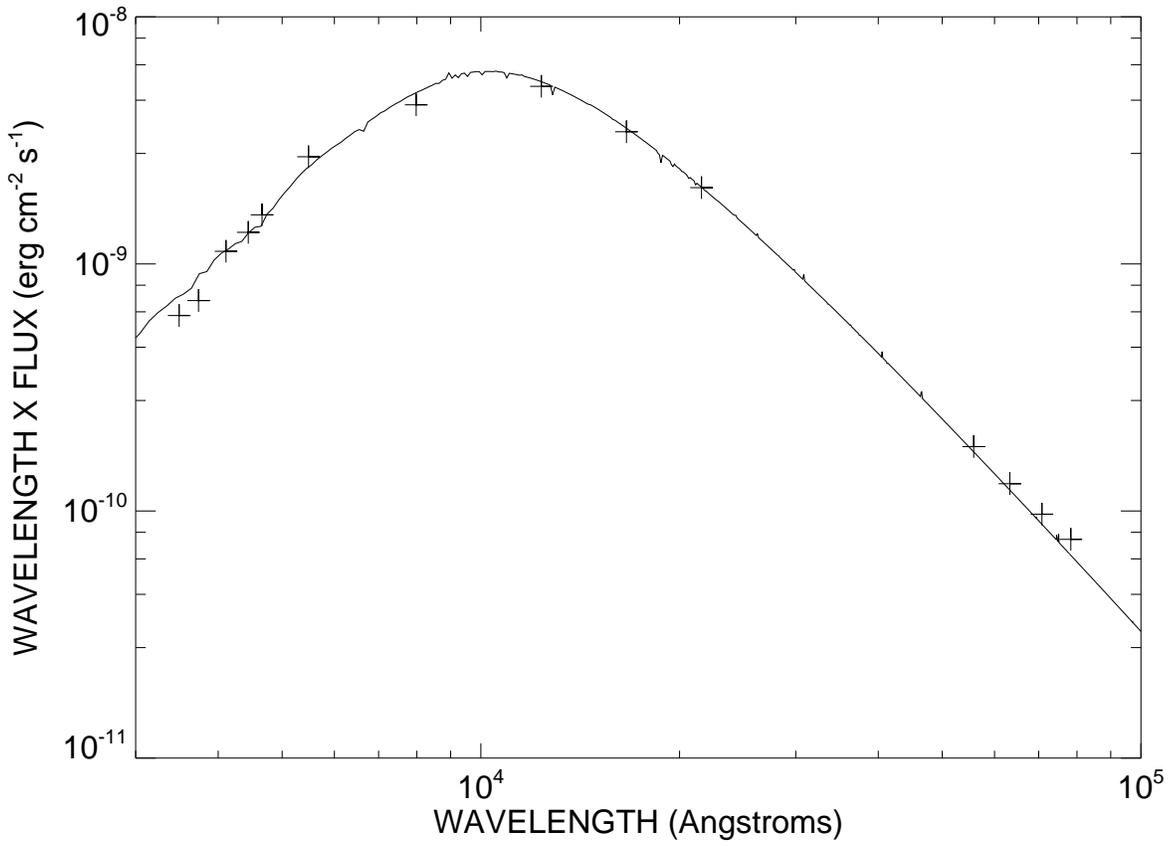}}
\end{center}
\caption{The spectral energy distribution for BD$+36^\circ 4063$. 
The plus signs indicate the observed fluxes while the solid line 
shows a reddened version of a BSTAR2006 flux model for the ON star.}
\label{fig2}
\end{figure}

\placefigure{fig2}     

We summarize the available constraints on the masses of the 
stars in Figure~3.  The mass ratio range 
derived from $V\sin i / K$ is indicated
by the lines of constant slope.  Next, we can use
the observed light curve to constrain the orbital inclination. 
The ON star probably fills out its Roche surface, and consequently 
the photometric variations are due to the star's
tidal distortion (geometric shape and gravity darkening). 
The amplitude of the ellipsoidal light curve depends on the 
degree of Roche filling (assumed complete) and the orbital 
inclination (larger at higher inclination).   Ideally we 
would create a model that includes both a Roche distorted mass 
donor and a disk surrounding the gainer \citep{dju08}, but 
given the small flux contribution of the disk and the relatively
large errors in the photometric data, we made an approximate 
model assuming that the companion is a small spherical object 
(with the same temperature as the ON star) 
that acts only as a mild flux dilution source in the light curve
of the tidally distorted ON star.  We used the GENSYN binary 
code \citep{moc72} to create model $I_C$ light curves 
for a grid of orbital inclinations, assuming that the ON 
star fills its Roche surface and that the companion has a 
radius that yields a monochromatic flux ratio, $F_2/F_1 = 0.1$ (matching that
of the preliminary tomographic reconstruction).
We obtained a best fit light curve (full amplitude $\approx 0.13$ mag) 
for an inclination $i= 49^\circ \pm 8^\circ$.  The derived masses based 
on the spectroscopic mass function and this range in inclination are 
sketched as curved lines in Figure~3. 

\begin{figure}
\begin{center}
{\includegraphics[angle=90,height=12cm]{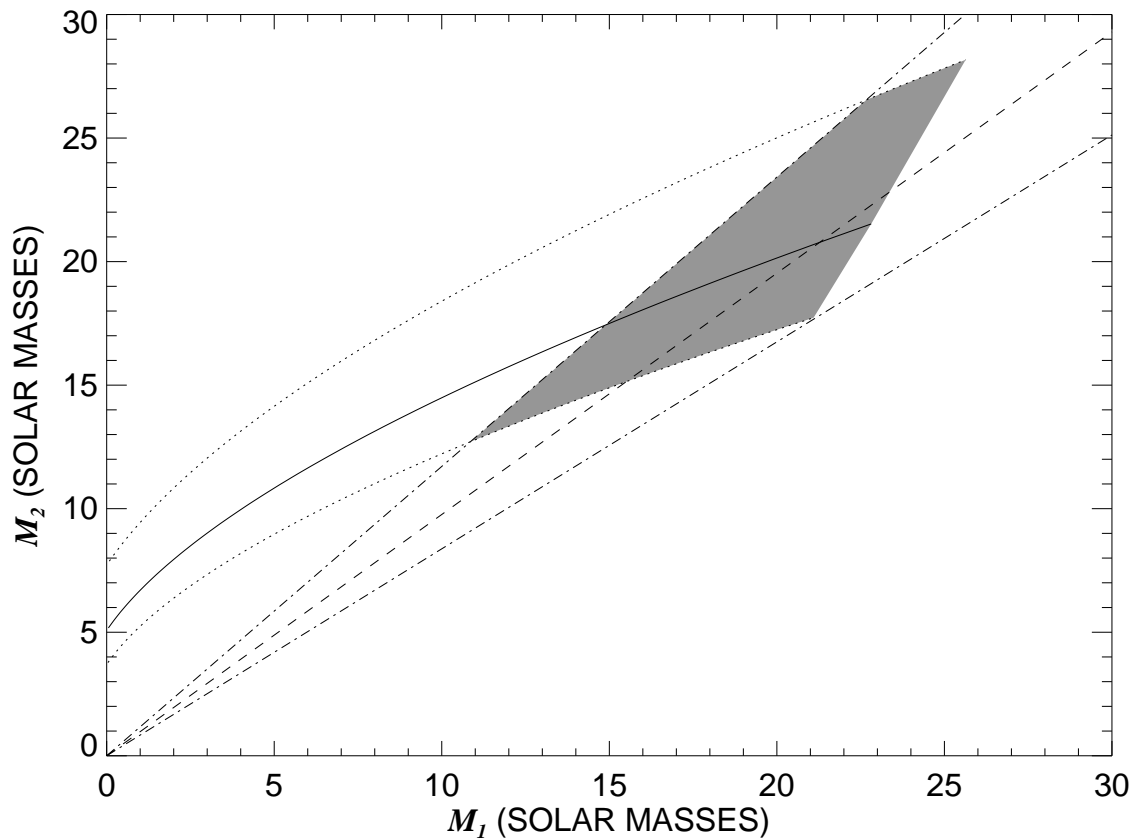}}
\end{center}
\caption{A mass plane diagram showing observational constraints on 
the masses of the ON star $M_1$ and the unseen companion $M_2$.
The dashed line shows the mass ratio derived from $V\sin i / K$ 
and the adjacent dot-dashed lines give the associated $\pm 1\sigma$ error 
range.  The solid line shows the relation from the spectroscopic 
mass function for an orbital inclination of $i=49^\circ$ derived from 
the light curve, while the dotted lines show the same for 
$i = 41^\circ$ ({\it above}) and $i = 57^\circ$ ({\it below}).
These three lines are terminated where the ON star would be underluminous
for its mass.  The shaded region indicates the ranges meeting all 
the constraints.}
\label{fig3}
\end{figure}

\placefigure{fig3}     

We can combine the radius -- distance relation derived from 
the SED with the projected rotational velocity (assuming 
a Roche filling, synchronously rotating, ON star) to form 
a distance -- inclination relation, 
$\sin i = (1.15 \pm 0.16) / d({\rm kpc}).$
The inclination range shown in Figure~3 corresponds
to distance range of 1.37 kpc ($i=57^\circ$) to 
1.76 kpc ($i=41^\circ$), which agrees well with 
distance estimates for the Cyg~OB1 association, 
$d = 1.25 - 1.83$~kpc \citep{uya01}.  We can also 
associate a stellar luminosity with any specific 
distance (or inclination) from the radius -- distance 
relation.  In the absence of mass transfer, we would expect the 
luminosity to agree with the mass -- luminosity relation 
for single stars of its temperature \citep{sch92}, 
so that one position along the constant inclination locus
would be preferred.  However, stars in binaries that 
suffer mass loss may appear overluminous for their mass, 
so in practice there is an upper limit along 
the inclination track where the star becomes more massive
than expected for the luminosity.  These terminal points
are indicated in Figure~3. Because the \citet{sch92} models to not 
account for rotation and because rapidly rotating stars may appear 
more luminous \citep{eks08}, these end-point values should be taken
as upper limits. The gray shaded region shows
the preferred mass ranges that fulfill all the constraints, and 
the optimal fit occurs for $M_1 \approx M_2 \approx 21 M_\odot$. 

The tentative picture that emerges from the spectroscopic analysis
is that the ON star is transferring gas to a relatively faint disk 
that surrounds and obscures the companion star.  The ON star 
must be very close to filling its Roche surface since we observe
evidence of its tidal distortion in the light curve and 
in the deepening of lines at conjunctions.  
The emission features that follow the orbital radial velocity 
characteristics of the hidden mass gainer probably form 
in the thick disk.  Thus, BD$+36^\circ 4063$ shares many features in
common with the W~Ser class of interacting binaries, but since the 
orbital period is much shorter and the mass ratio closer 
to unity than that found in other W~Ser systems, we 
suggest that BD$+36^\circ 4063$ represents an earlier and 
faster mass transfer stage of evolution.  If the system 
is coeval with the stars of Cyg~OB1 (with an age of 
approximately 7.5 Myr; \citealt{uya01}), then the mass -- 
radius -- age relations for single star evolutionary tracks 
\citep{sch92} suggest that the ON star probably began 
life with a mass of $25 M_\odot$ or less.  Thus, the ON star 
has probably lost only a modest fraction of its original 
mass so far.   BD$+36^\circ 4063$ offers us 
an important glimpse of binary evolution at its  
most intense stage, and we encourage new observational 
efforts to probe the system and its environment.  


\acknowledgments

We wish to thank the anonymous referee whose comments and suggestions
greatly helped clarify and strengthen our arguments.
We thank Dianne Harmer and the KPNO staff for their assistance
in making these observations possible.  We also thank 
Justin Howell (IPAC, Caltech) for help in interpreting the 
Spitzer data. This material is based on work supported by the
National Science Foundation under Grant AST-0606861.
This work is based in part on observations made with the
Spitzer Space Telescope, which is operated by the Jet Propulsion 
Laboratory, California Institute of Technology under a contract 
with NASA.




\begin{deluxetable}{lcccc}
\tablewidth{0pt}
\tablenum{1}
\tablecaption{Radial Velocity Measurements\label{tab1}}
\tablehead{
\colhead{HJD}           &
\colhead{Orbital}       &
\colhead{$V_{r}$}	&
\colhead{$\Delta V_{r}$}&
\colhead{$O-C$}         \\
\colhead{($-$2,400,000)}&
\colhead{Phase}         &
\colhead{(km s$^{-1}$)} &
\colhead{(km s$^{-1}$)} &
\colhead{(km s$^{-1}$)} }
\startdata
 54785.624 &  0.660 & \phs    126.7 &    2.4 & \phn\phs5.1 \\
 54786.576 &  0.858 & \phs    109.7 &    2.3 & \phn $-$0.8 \\
 54787.557 &  0.062 & \phn  $-$61.9 &    1.9 & \phs   16.7 \\
 54788.558 &  0.270 &      $-$180.0 &    2.2 & \phn $-$1.2 \\
 54789.567 &  0.480 & \phn  $-$36.7 &    1.8 & \phn\phs0.7 \\
 54790.556 &  0.685 & \phs    132.1 &    2.1 & \phn $-$1.3 \\
 54791.559 &  0.894 & \phs\phn 77.9 &    2.1 & \phn $-$6.8 \\
\enddata
\end{deluxetable}



\clearpage

\begin{thebibliography}{}

 \bibitem[Bagnuolo et al.(1994)]{bag94}
    Bagnuolo, W.~G., Jr., Gies, D.~R., Hahula, M.~E., Wiemker,~R., 
    \& Wiggs, M.~S. 
 1994, \apj, 423, 446
 \bibitem[Bessell et al.(1998)]{bes98}
    Bessell, M. S., Castelli, F., \& Plez, B. 1998, \aap, 333, 231
 \bibitem[Bolton \& Rogers(1978)]{bol78}
    Bolton, C.~T., \& Rogers, G.~L. 1978, \apj, 222, 234
 \bibitem[Cohen et al.(2003)]{coh03}
    Cohen, M., Wheaton, W. A., \& Megeath, S. T. 2003, \aj, 126, 1090
 \bibitem[Colina et al.(1996)]{col96}
    Colina, L., Bohlin, R., \& Castelli, F. 1996, HST Instrument Science
    Report CAL/SCS-008 (Baltimore: STScI)
 \bibitem[Crawford(1975)]{cra75}
    Crawford, D. L. 1975, \pasp, 87, 481
 \bibitem[Djura\v{s}evi\'{c} et al.(2008)]{dju08}
    Djura\v{s}evi\'{c}, G., Vince, I., \& Atanackovi\'{c}, O.
    2008, \aj, 136, 767
 \bibitem[Droege et al.(2006)]{dro06}
    Droege, T. F., Richmond, M. W., \& Sallman, M.
    2006, PASP, 118, 1666
 \bibitem[Ekstr{\"o}m et al.(2008)]{eks08} Ekstr{\"o}m, S., Meynet, G., 
    Maeder, A., \& Barblan, F.\ 2008, \aap, 478, 467
 \bibitem[Eggleton(1983)]{egg83}
    Eggleton, P.~P.  1983, \apj, 268, 368
 \bibitem[Fitzpatrick(1999)]{fit99}
    Fitzpatrick, E. L. 1999, \pasp, 111, 63
 \bibitem[Gies \& Bolton(1986)]{gie86}
    Gies, D.~R., \& Bolton, C.~T. 1986, \apj, 304, 371
 \bibitem[Gray(2005)]{gra05}
    Gray, D.~F. 2005, The Observation and Analysis of Stellar Photospheres 
    (3rd ed.; Cambridge: Cambridge Univ. Press)
 \bibitem[Gray(1998)]{gra98}
    Gray, R. O. 1998, AJ, 116, 482
 \bibitem[Grundstrom et al.(2007)]{gru07}
    Grundstrom, E. D.. Gies, D. R., Hillwig, T. C., McSwain, M. V., 
    Smith, N., Gehrz, R. D., Stahl, O., \& Kaufer, A, 2007, \apj, 667, 505
 \bibitem[Hanson et al.(1996)]{han96} Hanson, M.~M., Conti, 
    P.~S., \& Rieke, M.~J.\ 1996, \apjs, 107, 281
 \bibitem[Harries et al.(2002)]{har02}
    Harries, T. J., Howarth, I. D., \& Evans, C. J.
    2002, \mnras, 337, 341
 \bibitem[Hillier \& Miller(1998)]{hil98} Hillier, D.~J., 
    \& Miller, D.~L.\ 1998, \apj, 496, 407
 \bibitem[Hiltner(1956)]{hil56}
    Hiltner, W. A. 1956, \apjs, 2, 389
 \bibitem[Houck et al.(2004)]{hou04}
    Houck, J., et al. 2004, \apjs, 154, 18
 \bibitem[Humphreys(1978)]{hum78}
    Humphreys, R. M. 1978, \apjs, 38, 309
 \bibitem[Langer et al.(2008)]{lan08}
    Langer, N., Cantiello, M., Yoon, S.-C., Hunter, I., Brott, I., Lennon, D.,
    de Mink, S., \& Verheijdt, M. 2008, in 
    Massive Stars as Cosmic Engines, Proc. IAU, Vol. 3, Symp. S250,
    ed. F. Bresolin, P. A. Crowther, \& J. Puls (Cambridge: Cambridge Univ. 
    Press), 167
 \bibitem[Lanz \& Hubeny(2007)]{lan07}
    Lanz, T., \& Hubeny, I. 2007, \apjs, 169, 83
 \bibitem[Martins et al.(2005)Martins, Schaerer, \& Hillier]{mar05}
    Martins, F., Schaerer, D., \& Hillier, D.~J. 2005, \aap, 436, 1049	
 \bibitem[Mathys(1989)]{mat89}
    Mathys, G. 1989, \aaps, 81, 237
 \bibitem[Mochnacki \& Doughty(1972)]{moc72}
    Mochnacki, S. W., \& Doughty, N. A. 1972, \mnras, 156, 51
 \bibitem[Morbey \& Brosterhus(1974)]{mor74}
    Morbey, C., \& Brosterhus, E. B. 1974, \pasp, 86, 455
 \bibitem[Nazarenko \& Glazunova(2006)]{naz06}
    Nazarenko, V.~V., \& Glazunova, L.~V. 2006, Astr. Rep., 50, 369
 \bibitem[Patriarchi et al.(2003)]{pat03}
    Patriarchi, P., Morbidelli, L., \& Perinotto,  M. 2003, \aap, 410, 905
 \bibitem[Plavec(1980)]{pla80}
    Plavec, M. J. 1980, in Close Binary Stars: Observations and Interpretation
    (IAU Symp. 88), ed. M. J. Plavec, D. M. Popper, \& R. K. Ulrich
    (Dordrecht: Reidel), 251
 \bibitem[Ryans et al.(2002)]{rya02} 
    Ryans, R.~S.~I., Dufton, P.~L., Rolleston, W.~R.~J., Lennon, D.~J.,
    Keenan, F.~P., Smoker, J.~V., \& Lambert, D.~L. 2002, \mnras, 336, 577
 \bibitem[Schaller et al.(1992)]{sch92}
    Schaller, G., Schaerer, D., Meynet, G., \& Maeder, A. 1992, \aaps, 96, 269
 \bibitem[Skrutskie et al.(2006)]{skr06}
    Skrutskie, M. F., et al. 2006, \aj, 131, 1163
 \bibitem[Smith et al.(2002)]{smi02} 
    Smith, N., Gehrz, R.~D., Stahl, O., Balick, B., \& Kaufer, A. 2002, 
    \apj, 578, 464
\bibitem[Tamblyn et al.(1996)]{tam96} Tamblyn, P., Rieke, 
    G.~H., Hanson, M.~M., Close, L.~M., McCarthy, D.~W., Jr., 
    \& Rieke, M.~J.\ 1996, \apj, 456, 206 
 \bibitem[Tarasov(2000)]{tar00}
    Tarasov, A.~E. 2000, in The Be Phenomenon in Early-Type Stars 
    (ASP Conf. Ser. 214), ed. M.~A. Smith, H.~F. Henrichs, \& J. Fabregat 
    (San Francisco: ASP), 644
 \bibitem[Uyaniker et al.(2001)]{uya01}
    Uyaniker, B., F\"{u}rst, E., Reich, W., Aschenbach, B., \& Wielebinski, R.
    2001, \aap, 371, 675
 \bibitem[Wade \& Rucinski(1985)]{wad85}
    Wade, R.~A., \& Rucinski, S.~M. 1985, \aaps, 60, 471
 \bibitem[Walborn \& Howarth(2000)]{wal00}
    Walborn, N.~R., \& Howarth, I. D. 2000, \pasp, 112, 1446
 \bibitem[Wo\'{z}niak et al.(2004)]{woz04}
    Wo\'{z}niak, P. R., et al. 2004, AJ, 127, 2436
\bibitem[Wellstein et al.(2001)]{wel01} Wellstein, S., 
    Langer, N., \& Braun, H.\ 2001, \aap, 369, 939
 \bibitem[Zhao et al.(2008)]{zha08}
    Zhao, M., et al. 2008, \apj, 684, L95
\end{thebibliography}
\end{document}